\documentclass[showpacs,preprintnumbers,amsmath,amssymb,prd]{revtex4}
\usepackage{epsfig}
\usepackage{graphicx}
\usepackage{dcolumn}
\usepackage{bm}
\begin{document}
\title{\boldmath 
Precise measurement of spin-averaged $\chi_{cJ}(1P)$ mass using photon conversions
in $\psi(2S) \rightarrow \gamma \chi_{cJ}$
}
\author{
M.~Ablikim$^{1}$,              J.~Z.~Bai$^{1}$,               Y.~Ban$^{11}$,
J.~G.~Bian$^{1}$,              X.~Cai$^{1}$,                  J.~F.~Chang$^{1}$,
H.~F.~Chen$^{17}$,             H.~S.~Chen$^{1}$,              H.~X.~Chen$^{1}$,
J.~C.~Chen$^{1}$
% ...
\\(BES Collaboration)\\
}
\affiliation{
$^1$ Institute of High Energy Physics, Beijing 100049, People's Republic of China \\
% ...
$^{11}$ Peking University, Beijing 100871, People's Republic of China\\
% ...
$^{17}$ University of Science and Technology of China, Hefei 230026, People's Republic of China\\
% ...
}
\begin{abstract}
Using photon conversions to $e^+e^-$ pairs,
the energy spectrum of inclusive
photons from $\psi(2S)$ radiative decays is measured
by BESII at the Beijing Electron-Positron Collider.
The $\chi_{cJ}(1P)$ states (J=0,1,2) are clearly observed
with energy resolution between 2.3 to 3.8 $\hbox{MeV}$,
and their masses and the spin-averaged $\chi_{cJ}$ mass are
determined to be $M_{\chi_{c0}}=3414.21\pm 0.39\pm 0.27$,
$M_{\chi_{c1}}=3510.30\pm 0.14\pm 0.16$, $M_{\chi_{c2}}=3555.70\pm
0.59\pm 0.39$ and
$M(^3P_{cog})=3524.85\pm 0.32\pm 0.30$ $\hbox{MeV}/c^2$. 
\end{abstract}
\maketitle

\section{Introduction}
Precise measurements of the spectrum and the decay properties of
charmonia are essential to test Potential QCD models and QCD based
approaches~\cite{eichten}.  There is renewed interest since the
discovery of the X(3872)~\cite{X3872} and the observations of the
expected $\eta_C(2S)$ and $h_C$ $(^1P_1)$ states~\cite{etac and hc},
and there has been recent progress, both theoretically and
experimentally~\cite{QCD review}.  There are more accurate
determinations of the charmonium mass spectrum and radiative
transition rates using both a relativistic quark model with
relativistic corrections of order $v^2/c^2$~\cite{ebert} and a
potential model with a semirelativistic approach~\cite{Radford}.  The
$\psi(2S)$ mass and width have been redetermined with an updated
radiative correction~\cite{artamonov}, and newly measured with better
precision~\cite{Aulchenko}. In addition to previous measurements of
$\chi_{cJ}$ states~\cite{pdg04}, two $\chi_{c0}$ measurements by
E835~\cite{bagnosco} and new $\chi_{cJ}$ (J=0,1,2) measurements by
CLEO~\cite{CLEO chicJ} have been recently published.  Improved
precision on $\chi_{cJ}$ masses is important for the determination of
the singlet-triplet splitting, $M(^1P_1)-M(^3P_{cog})$,
which is predicted by lattice QCD and nonrelativistic QCD
~\cite{Godfrey and Rosner}. Here $M(^3P_{cog})$ is the spin-averaged
$^3P_J$ mass for the $\chi_{cJ}$ states ($J=0,1,2$).

In this paper, results on the $\chi_{cJ}$ masses (J=0,1,2) and widths
(J=0,1) from a measurement of the energy spectrum of inclusive photons
in $\psi(2S)$ radiative decays, using photon conversions to improve the
energy resolution, are presented.  The measurement uses $14\times
10^6$ $\psi(2S)$ events collected with the upgraded Beijing
Spectrometer (BESII) at the BEPC Collider.

\section{BES Detector and Monte Carlo simulation}
The BESII detector is described elsewhere~\cite{bes2}. A 12-layer
vertex chamber (VC) surrounding the beam pipe provides 
hit information in trigger criteria for charged tracks. Charged particle
momenta are determined with a resolution of $\sigma _p/p = 1.78\%
\sqrt{1+p^2}$ ($p$ in $\hbox{\rm GeV}/c$) in a 40-layer cylindrical
drift chamber (MDC).  Particle identification is accomplished 
by measurements of ionization ($dE/dx$) in the MDC and
time-of-flight (TOF) in a barrel-like array of 48
scintillation counters. The $dE/dx$ resolution is $\sigma_{dE/dx}=
8\%$; the TOF resolution is $\sigma_{TOF}$=200 ps for hadrons.  A
12-radiation-length barrel shower counter (BSC) measures energies of
photons with a resolution of $\sigma_E/E=21\%/\sqrt{E}$ ($E$ in
GeV). A solenoidal coil supplies a 0.4~Tesla magnetic field over the
tracking volume.

A Geant3 based Monte Carlo (MC) SIMBES~\cite{liuhm}, which simulates
the detector response including interactions of secondary particles in
the detector material, is used to determine the energy resolution and
detection efficiency of photons reconstructed from their converted
$e^+e^-$ pairs, as well as to optimize selection criteria and estimate
backgrounds.  Under the assumption of a pure E1 transition for the
$\psi(2S)\rightarrow \gamma\chi_{cJ}$ decays, the polar angle
($\theta$) distributions of the photons are given by $1+k\cos^2\theta$
with $k = 1,-\frac{1}{3},\frac{1}{13}$ for $J=0,1,2$,
respectively~\cite{karl}.

Good energy resolution for low energy photons is essential for precise
measurements of $\chi_{cJ}$ masses and widths from fitting the
photon spectrum of $\psi(2S)$ radiative decays.  Momentum resolution
of about 1.6 to 4.1 $\hbox{MeV}/c$ can be obtained for low momentum electrons
from 60 to 250 $\hbox{MeV}/c$.  Photons from
$\psi(2S)\rightarrow \gamma \chi_{cJ}$ decays have energies of about
261, 171, and 128 $\hbox{MeV}$ for the $\chi_{cJ}$ final states
(J=0,1,2), and the electrons produced in photon conversions
occur in this low momentum region.

\section{Photon reconstruction and selection}
\begin{figure}[h]
\centerline{
\hbox{\psfig{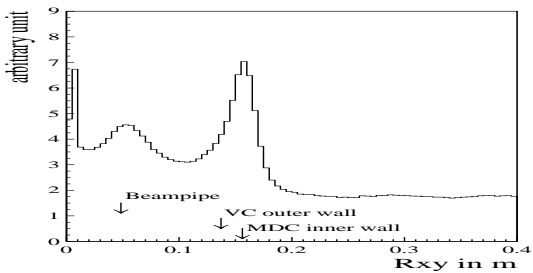}}}
\caption{The $R_{xy}$ distribution for gamma
conversions to $e^+e^-$ pairs in the BESII detector from hadron 
events in the 58 $\times 10^6$ $J/\psi$ event sample.}
\label{rxy}
\end{figure}

We choose two oppositely charged tracks with each track having a good
helix fit and a polar angle with $|\cos\theta|< 0.8$.  The
intersection of the electron and positron trajectories in the xy-plane
(the beam line is the $z$ axis) is determined, and this point is taken as
the photon conversion point (CP).  The photon conversion length
$R_{xy}$ is defined as the distance from the beam line to the CP in
the xy-plane.  Fig.~\ref{rxy} shows the $R_{xy}$ distribution for
photon conversions to $e^+e^-$ pairs in the BESII detector for
hadronic events in the 58 $\times 10^6$ $J/\psi$ event sample.  The two
broad peaks in Fig.~\ref{rxy} correspond to the beampipe region, where the
beampipe, the VC, and inner wall of the MDC are located. Combinatorial
background from charged hadron tracks is also seen in the
$R_{xy}<2$ cm region. Equivalent materials in the beampipe wall, VC,
VC outer wall, and the MDC inner wall are $0.536$, $0.657$,$0.375$, and
$1.107$ in units of $0.01X_0$~\cite{hongt}, respectively, where $X_0$
is a radiation length.  The electron and positron directions are
calculated at the photon conversion point, and their momenta are
corrected to that point.

\begin{figure}[htbp]
\centerline{
\hbox{\psfig{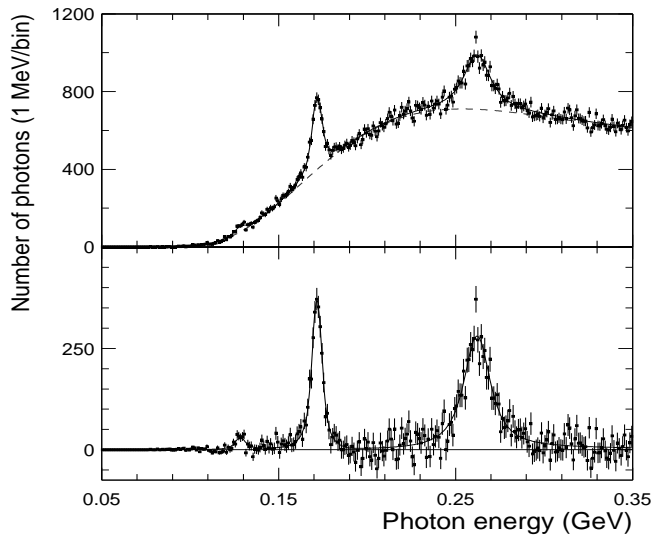}}}
\caption{Inclusive photon spectrum from photon conversions from $14
  \times 10^6$ $\psi(2S)$ events. A fit (described in the text) is
  made to $\psi(2S) \rightarrow \gamma \chi_{cJ}$ decays $(J=0,1,2)$
  plus threshold background. Points with error bars are data. The
  solid line is the fit; the dashed line is the fitted
  background. Background subtracted results are shown in the lower
  plot.}
\label{plot fit data}
\end{figure}

Good photons are selected.  The photon conversion length must lie
within the beampipe region, $2 < R_{xy}< 22$ cm, and the invariant mass
of an $e^+e^-$ pair is required to satisfy $M_{e^+e^-} < 20$
$\hbox{MeV}/c^2$. Combinatorial background from charged hadron tracks
is further removed by requiring $\cos\theta_{defl} > 0.9$ , where
$\theta_{defl}$ is the deflection angle between the photon momentum
and photon track (a vector from the beam to the CP). To suppress
background from beam-gas and beam-pipe interactions, the total energy
in the event must satisfy $E_{tot} > E_{beam}/2$ and momentum
asymmetry must satisfy $dp_{asym} < 0.9$.
Here $dp_{asym}$ is defined
as a ratio of the vector sum to the scalar sum of the momenta of all charged
and neutral tracks in the event.  The observed photon energy spectrum
from the $\psi(2S)$ data after the selection of good photons is shown
in Fig.~\ref{plot fit data}.  The spectrum shows the $\chi_{cJ}$
states plus a large background. The sharp drop at low energy is mainly
caused by low photon detection efficiency.

\section{$dE/dx$ correction and photon energy scale}
Energy loss dE/dx by ionization for electrons traversing a small
thickness of material with energy above a few tens of $\hbox{MeV}$ can
be described by the Bethe-Bloch equation~\cite{pdg04}.  The dE/dx
correction for charged particles, produced near the beamline and
traversing the whole beampipe region, should take into account the full
thickness of material in the region. However, $e^+e^-$
pairs from photon conversions are mostly produced in the region where the
VC outer wall and the MDC inner wall are located.  Thus the effective
thickness of material between the location, where a pair is produced, and
first layer of the MDC wires must be estimated for each
electron pair.  The procedure to make $dE/dx$ corrections for
electrons has two steps : (1) A preliminary $dE/dx$
correction using half the full
thickness of all the materials in the beampipe region is made. Good
photons are reconstructed, and their conversion lengths $R_{xy}$ are
calculated.  (2) The final $dE/dx$ corrections are estimated based on
these $R_{xy}$.

\begin{figure}[htbp]
\centerline{
\hbox{\psfig{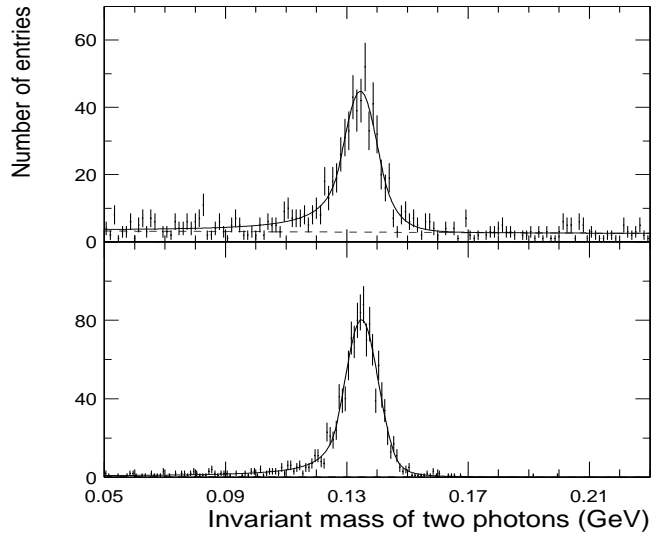}}}
\caption{Invariant mass distribution of photon pair from $J/\psi$ data (top)
and Monte Carlo (bottom) events. The solid line is the fitted curve for
signal plus background. The dashed line is the fitted curve for background.
}\label{plot fit pi0 mass}
\end{figure}

The energy scale of photons reconstructed from $e^+e^-$ pairs is
studied using simulated MC events and data. $63\times 10^6$ $\pi^0$ 
signal events are generated using MC technique, with 
same momentum and polar angular distributions as that found 
from $\pi^0$ data sample. A sample of $\pi^0$
mesons decaying to two photons with both photons converting to $e^+e^-$
pairs is selected from $58\times 10^6$ $J/\psi$ events.  To suppress
hadron contamination, electron identification is required and good
photons are selected. Background is further suppressed with
additional requirements on the photon energy, $E_{\gamma} \leq 1GeV$,
and the opening angle between the two photons,
$0.75<|\cos\theta_{\gamma\gamma}|<0.97$.  The invariant mass
distribution of two photons for both MC and data, after the specific 
$dE/dx$ correction for electrons
described above, is fitted with the improved Crystal Ball (ICB)
function 
(defined in section~\ref{Physics problem}) plus a first order 
polynomial background. The results are shown in Fig.~\ref{plot fit pi0 mass}. 
The resulting $\pi^0$ masses 
$(134.47\pm 0.42)$ $\hbox{MeV}/c^2$ in data and
$(134.86\pm 0.20)$ $\hbox{MeV}/c^2$ in MC are
consistent with the PDG value of $134.98$ $\hbox{MeV}/c^2$~\cite{pdg04}.
The corresponding mass resolutions $(5.70\pm 0.58)$ $\hbox{MeV}/c^2$ 
in data and $(5.55\pm 0.21)$ $\hbox{MeV}/c^2$ in MC agree
within errors. The $\chi^2$/D.F. (degree of freedom) for the fits
are 126/103 in data and 117/140 in MC.

\section{Physics problem and detector resolution}\label{Physics problem}
The energy of a photon from $\psi(2S)\rightarrow 
\gamma\chi_{cJ}$ decay is given by
\begin{eqnarray}
E_\gamma &=& (M^2_{\psi(2S)}-M^2_{\chi_{cJ}}) / 2M_{\psi(2S)}, 
\label{energyg}
\end{eqnarray}
where $M_{\psi(2S)}$ and $M_{\chi_{cJ}}$ are the masses of the
$\psi(2S)$ and $\chi_{cJ}$, respectively. 
The $\psi(2S)$  and $\chi_{cJ}$ widths must 
be taken into account;  $M_{\psi(2S)}$ 
and $M_{\chi_{cJ}}$ are described by Breit-Wigner functions (2D problem). 
By taking $x={M_{\psi(2S)}}$, the probability density function (pdf) for the 
photon energy $E_\gamma$ can be written as~\cite{mfisz}
\begin{eqnarray}
f_{pdf}(E_\gamma) = \int BW(x) BW(M_{\chi_{cJ}})
\frac{x}{M_{\chi_{cJ}}} dx,  
\label{egpdf2}
\end{eqnarray}
where $M_{\chi_{cJ}}$ depends on $E_\gamma$ by Eq.~(\ref{energyg}).

\begin{figure}[htbp]
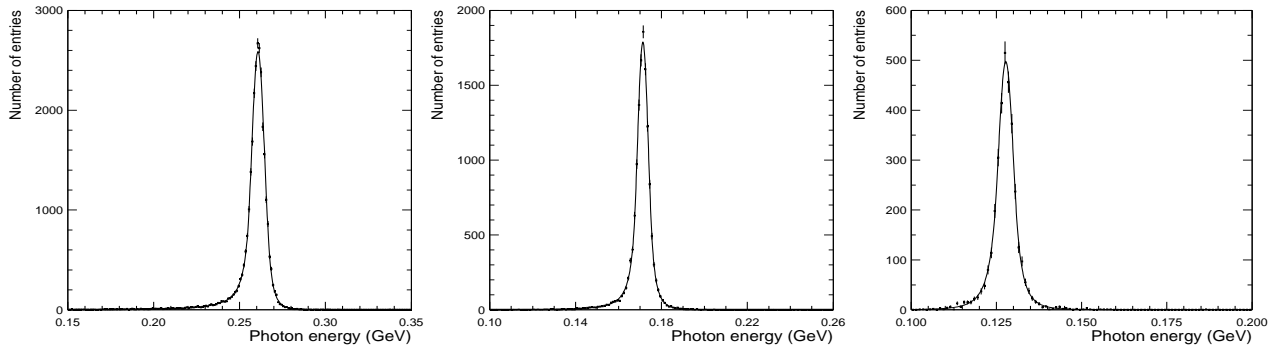

\centerline{
\hbox{\psfig{file=plot_c0w0_icb_mc.epsi,height=4.5cm,width=5.5cm}}
\hbox{\psfig{file=plot_c1w0_icb_mc.epsi,height=4.5cm,width=5.5cm}}
\hbox{\psfig{file=plot_c2w0_icb_mc.epsi,height=4.5cm,width=5.5cm}}}
\caption{Energy distributions of signal photons from
$\psi(2S) \rightarrow \gamma\chi_{cJ}$ decays with zero widths of both
$\psi(2S)$ and $\chi_{cJ}$ for $\chi_{c0}$(left), $\chi_{c1}$(mid)
and $\chi_{c2}$(right) final states fitted to the ICB function. 
The points are MC data. The solid line are the fitted curve.
}\label{photon resolution}
\end{figure}

As a result of traversing material in the beam pipe region, the
electron energy is smeared due to energy loss by ionization, and a
long tail on the low side of the energy distribution is induced by
bremsstrahlung radiation.  Multiple scattering of electrons,
especially at large angles, gives tails on both sides of the photon
energy distribution of photon conversions.  The photon energy
resolution from photon conversions can be nicely modeled by our Geant3
MC simulation, and well fitted by the ICB function. The original 
Crystal Ball (CB) function has a Gaussian in its central and upper 
energy region but long tail at lower energy region~\cite{mnfit}. 
The improved CB function is defined as same as the CB function but has 
an additional tail at its upper side. The photon energy
distributions from large MC samples of $\psi(2S) \rightarrow
\gamma\chi_{cJ}$ decays $(J=0,1,2)$, with zero widths for both the
$\psi(2S)$ and $\chi_{cJ}$ states, are fitted to ICB functions and
shown in Fig.~\ref{photon resolution}. 
The $\chi^2$/D.F. from the fits are 103.8/93, 37.7/53 and 47.6/43
corresponding to $\chi_{c0}$, $\chi_{c1}$, $\chi_{c2}$ states. Five
parameters in the ICB function, the photon energy resolution
and four empirical parameters to describe the
tails on the lower and upper sides are determined from the fits and
used as input
parameters in the detector resolution function for each decay
mode. Photon energy resolutions for the $\psi(2S) \rightarrow
\gamma\chi_{cJ}$ decays (J=0,1,2) are found to be $3.78\pm 0.04$,
$2.58\pm 0.05$, and $2.26\pm 0.11$ $\hbox{MeV}$, respectively.

The energy dependencies of the photon detection efficiency and resolution
are determined using MC simulation in the energy range between 100 and
400 (300, 220) $\hbox{MeV}$ for $\psi(2S) \rightarrow
\gamma\chi_{c0}~(\gamma\chi_{c1},\gamma\chi_{c2})$ decays. 
The efficiency includes the
effects of detector geometry, MDC tracking, photon
reconstruction, and the spin-dependent $\cos\theta$ distribution.

\section{Fitting three photon spectrum}
Background contamination from
$\chi_{c0} \rightarrow \gamma J/\psi$ decay is negligible due to its
small branching fraction.  To exclude photons from the $\chi_{c1,c2}
\rightarrow \gamma J/\psi$ decays, the photon energy range in the fits is
chosen to be within $90~\hbox{MeV} \le E_\gamma \le 350~\hbox{MeV}$.
An enhancement from the decay $\psi(2S) \rightarrow \eta J/\psi$ with
$\eta\rightarrow \gamma\gamma$ in a region of $180$ to
$400~\hbox{MeV}$ is estimated by MC method, and subtracted from the
data, according to the measured total number of
$\psi(2S)$ events and known branching fractions.
The smooth background under the signal photon lines can be described
by a threshold function 
from the $Mn\_Fit$ package~\cite{mnfit}.
\begin{eqnarray}
bg_{threshold}(x) &=& B\cdot (x-x_0)^{\alpha}\cdot e^{c_1(x-x_0)+c_2(x-x_0)^2},
\label{threshold fun}
\end{eqnarray}
where B, $x_0$, and $\alpha$ are normalization factor, threshold,
and power; $c_1$ and $c_2$ are coefficients linear and quadratic in x.
The threshold function has been used to fit backgound with a threshold at
the lower or upper side of an observed distribution by some 
experiments~\cite{Babar}.

Considering the physical photon energy $x\equiv E_\gamma$ 
and its error
$y \equiv \triangle E_\gamma$ due to detector resolution, the
measured photon energy is $u=x + y$ and its pdf function can be written
as :
\begin{eqnarray}
h_{pdf}(u) = \int c_{E_1}(u-y)\cdot c_{eff}(u-y)\cdot f_{pdf}(u-y)\cdot g_{res}(y)dy 
= \int c_{E_1}(x)\cdot c_{eff}(x)\cdot f_{pdf}(x)\cdot g_{res}(u-x)dx,
\label{pdf function of photon energy}
\end{eqnarray}
where $c_{E_1}(x) = x^3/E^3_{\gamma,\chi_{cJ}}$,
$c_{eff}(x)=\epsilon_{\chi_{cJ}}(x)/\epsilon(E_{\gamma,\chi_{cJ}})$,
$f_{pdf}(x)$ is defined with Eq.~(\ref{egpdf2}), and $g_{res}(y)$
is the ICB resolution function. 
With the assumption that the $E1$ electric dipole transition
for $\psi(2S) \rightarrow \gamma \chi_{cJ}$ decays (J=0,1,2) 
dominates, an $E^3_\gamma$ energy dependence 
is included in the folded signal shape. 
The detection efficiency 
$\epsilon_{\chi_{cJ}}(x)$ and energy
resolution as a function of photon energy are included in the
fitting. 
Normalization factors $E_{\gamma,\chi_{cJ}}$ and 
$\epsilon(E_{\gamma,\chi_{cJ}})$ are photon energy corresponding to 
fitted $\chi_{cJ}$ mass and efficiency at the photon energy, 
respectively. Notice that parameters, masses $M_{\chi_{cJ}}$ and
widths $\Gamma_{\chi_{cJ}}$~(J=0,1,2),
are implicitly contained in the $f_{pdf}(x)$ function, and 
detector resolution and tail parameters are in the $g_{res}(y)$ function.
The likelihood function, $Lk(u;M_{\chi_{cJ}},\Gamma_{\chi_{cJ}})$,
is constructed with three $\chi_{cJ}$ signals plus threshold background:
\begin{eqnarray}
Lk(u;M_{\chi_{cJ}},\Gamma_{\chi_{cJ}}) = bg_{threshold}(u) + 
\sum_{J=0}^{2} A_J\cdot h_{pdf,\chi_{cJ}}(u;M_{\chi_{cJ}},\Gamma_{\chi_{cJ}}).
\label{pdf function of photon energy}
\end{eqnarray}
Here $A_J$ is the observed area in each $\chi_{cJ}$ signal.

An input-output test is performed to verify the accuracy of the
fitting algorithm for the 2D problem using MC events.  The energy
dependencies of the photon detection efficiency and resolution are
included in the fitting procedure.  A sample of MC events for the
$\psi(2S) \rightarrow \gamma\chi_{cJ}$ decays with non-zero width for
both the $\psi(2S)$ and $\chi_{cJ}$ are produced. The photon energy
distribution is fitted with the 2D pdf function convoluted with the
ICB resolution.  The resulting masses and widths of the $\chi_{cJ}$
states in this test are consistent with the MC input parameters.

A combined fit of the three photon spectra corresponding to the
$\psi(2S)\rightarrow \gamma\chi_{c0}, \gamma\chi_{c1},
\gamma\chi_{c2}$ decays is performed to three 2D pdf functions (see
Eq.~(\ref{egpdf2})), each convoluted with its ICB resolution function,
plus threshold background. The $\chi_{c2}$ width is fixed in the fit
due to the limited statistics. The $\chi^2$ and D.F. (degrees of freedom)
from the fit are 521.0 and (520-13), where total number of free paramaters
is 13. The effect of the beam energy spread~\cite{beam energy
spread} in the measurement is also included, but is found to be
negligible due to the narrow width of the $\psi(2S)$ state.  A study
shows that the bin size ($0.5$ or $0.2~\hbox{MeV}$) in the binned fits
slightly affects the fitted masses and widths of the $\chi_{cJ}$
states. The difference in the results due to different bin sizes are
added to the systematic error. The results of the binned fit (0.5
$\hbox{MeV}/$bin) are shown in Fig.~\ref{plot fit data}. 

\section{Systematic errors}
Samples of QED radiative two photon events with one photon converting
to an $e^+e^-$ pair are selected for both data and MC simulation.  The
two photons are required to be emitted back-to-back.  The fitted
photon energy in data is different from the expected MC value by
$-1.2\sigma$ ($\sigma$~=~0.86 $\hbox{MeV}$), which has a relative
error at the same level as a correction factor $s = 0.9975\pm 0.0007$
for the magnetic field~\cite{field correction}. Thus a relative
precision of 0.0007 is added as the systematic error in the photon energy
determination.

The selection of $\pi^0\rightarrow\gamma\gamma$ events with both
photons converting to $e^+e^-$ pairs from $58\times 10^6$ $J/\psi$
events yields a data sample of $503$ $\pi^0$ mesons.  A MC sample of
$\pi^0$ mesons is generated with the same momentum and polar angular
distributions as found from the $\pi^0$ data sample.  The $\pi^0$
mass resolutions determined from the data and MC are in good
agreement; their difference is $0.15\pm 0.62~\hbox{MeV}/c^2$.  We
assume that the photon energy resolution and uncertainty in the
direction of the photon momentum each contribute half in the $\pi^0$
mass resolution.  Hence, the difference $\triangle\sigma_{M_{\pi^0}}$
of the $\pi^0$ mass resolutions between MC and data from the uncertainty of
the photon energy resolution lies within $(-0.29,
+0.59)~\hbox{MeV}/c^2$ with a probability of $68.3\%$.  We assume
$\triangle\sigma_{E_\gamma}(\chi_{cJ}) /
\sigma_{E_\gamma}(\chi_{cJ},MC) = \triangle\sigma_{M_{\pi^0}} /
\sigma_{M_\pi^0}(MC)$, where $\sigma_{E_\gamma}(\chi_{cJ},MC)$ and
$\triangle\sigma_{E_\gamma}(\chi_{cJ})$ are the MC photon energy
resolution and the difference between MC and data for $\chi_{cJ}$
final states, and $\sigma_{M_\pi^0}(MC)$ and $\triangle\sigma_{M_{\pi^0}}$
are $\pi^0$ mass resolution in MC and the difference between MC and
data. Thus $1\sigma$ confidence intervals of
$\triangle\sigma_{E_\gamma}(\chi_{cJ})$ for the
$\psi(2S)\rightarrow\gamma\chi_{c0},\gamma\chi_{c1}$ decays are
estimated to be $(-0.20,+0.40)$ and $(-0.14,+0.27)$ $\hbox{MeV}$,
which are used to estimate systematic errors in the determination of
the $\chi_{c0}$ and $\chi_{c1}$ widths.

The effect of the background shape uncertainty 
is studied using $\psi(2S)$ data and $\psi(2S)\rightarrow anything$
MC~\cite{chenjc}.  The relative differences in background shape
parameters between floated and fixed widths of the $\chi_{c0,c1}$
states are determined in fits for MC data, and fed back to correct
background parameters in the fits for data. 
The difference between results for $\psi(2S)$ data with the
background shape floated and fixed is taken 
as a systematic error.
In addition, our MC study with non-zero width of both $\psi(2S)$ and
$\chi_{cJ}$ shows that differences in the fitted masses from input
values for the $\chi_{c0}$ and $\chi_{c1}$ are $0.12\pm0.06$ and
$0.06\pm0.03~\hbox{MeV}/c^2$, while that for the $\chi_{c2}$
is as large as $0.31\pm0.06~\hbox{MeV}/c^2$.  The differences are
attributed to uncertainties in the energy loss correction for low momentum
electrons.  The systematic errors, including the contributions from
the uncertainties of the photon detection efficiency, are summarized in the
Table~\ref{table sys error}.
\begin{table}[htbp]
\caption{Summary of systematic errors in the determination of the 
$\chi_{cJ}$ masses and widths (in $\hbox{MeV}/c^2$).}
\begin{center}
\begin{tabular}{|lccccc|}
\hline
source  &$M_{\chi_{c0}}$&$\Gamma_{\chi_{c0}}$&$M_{\chi_{c1}}$&$\Gamma_{\chi_{c1}}$&$M_{\chi_{c2}}$ \\\hline
background shape &$0.03$   & $0.8$   & $0.02$ &  $0.07$ &  $0.04$               \\
correction in magnetic field &$0.19$   &         &$0.13$  &        &  $0.09$              \\
MC simulation in $\sigma_{E_\gamma}$ & &  $^{+0.29}_{-0.76}$ &&$^{+0.25}_{-0.77}$&    \\
different bin size & $0.02$  &$0.02$ &$0.01$     &$0.01$ & $0.02$     \\
photon energy correction   & $0.18$  &       &$0.09$     &       &  $0.37$     \\
efficiency uncertainty & $0.04$  & $0.03$&$0.01$     &$0.00$ &$0.04$       \\
error of $M_{\psi(2S)}$&$0.034$ & &$0.034$    &       &  $0.034$     \\\hline
total        & $0.27$  &$^{+0.85}_{-1.10}$ &$0.16$ &$^{+0.26}_{-0.77}$ & $0.39$     \\\hline
\end{tabular}
\end{center}
\label{table sys error}
\end{table}

\section{Results and discussion}
With good energy resolution for low energy photons obtained using
photon conversions, the precise measurement of the masses and widths of
$\chi_{cJ}~(J=0,1,2)$ states from inclusive $\psi(2S)$ radiative decays 
can be obtained. The masses and widths are determined to be
$M_{\chi_{c0}}=3414.21\pm 0.39\pm 0.27$, $M_{\chi_{c1}}=3510.30\pm
0.14\pm 0.16$, $M_{\chi_{c2}}=3555.70\pm 0.59\pm 0.39$
$\hbox{MeV}/c^2$,
$\Gamma_{\chi_{c0}}=12.6{^{+1.5}_{-1.6}}{^{+0.9}_{-1.1}}$ and
$\Gamma_{\chi_{c1}}=1.39{^{+0.40}_{-0.38}}{^{+0.26}_{-0.77}}$
$\hbox{MeV}/c^2$. The mass splittings in the $\chi_{cJ}(1P)$ triplet
and their ratio are found to be $\triangle M_{21} = M_{\chi_{c2}} -
M_{\chi_{c1}} = 45.40\pm0.61\pm0.42~\hbox{MeV}/c^2$, $\triangle M_{10}
= M_{\chi_{c1}} - M_{\chi_{c0}} = 96.09\pm0.41\pm0.31~\hbox{MeV}/c^2$
and $\rho (\chi_c) = \triangle M_{21} / \triangle M_{10} = 0.472\pm 0.006\pm
0.004$.
For the first time, 
the spin-averaged $^3P_J$ mass (weighted with the factors $2J+1$) 
for the $\chi_{cJ}$ states is precisely measured in one experiment and
determined to be $M(^3P_{cog})=3524.85\pm 0.32\pm 0.30$ $\hbox{MeV}/c^2$.
The first errors in the results are statistical and the
second are systematic. 
Correlations are taken into account in estimations of both statistical 
and systematic errors for the $\triangle M_{21}$, 
$\triangle M_{10}$, $\rho (\chi_c)$ and $M(^3P_{cog})$.
Correlation coefficients between mass parameters for statistical error
are obtained from the error matrix of the combined fit, and
that for systematic error are all assumed to be 1.

The $\chi_{cJ}$ masses (J=0,1,2) determined here are consistent with
the recent measurements by CLEO~\cite{CLEO chicJ}, but have smaller
systematic errors. The precisions for the $\chi_{c0}$ and $\chi_{c1}$
masses are compatible with those of previous measurements by
E835~\cite{bagnosco} and E760~\cite{pdg04}, while that for the
$\chi_{c2}$ mass is not as good as theirs due to low statistics. Note
that our $\chi_{c0}$ mass is lower than that measured by the E835 via
$\chi_{c0}\rightarrow \gamma J/\psi$ decay by $1.2~\hbox{MeV}$
(corresponding to $1.8\sigma$), but agrees with their later measurement
via $\chi_{c0}\rightarrow \pi^0\pi^0$ decay.  The width of the
$\chi_{cJ}$ states (J=0,1) determined here are also consistent
with their values; larger errors in our widths are caused by
limited statistics for both signal photons and inclusive $\pi^0$ mesons.

\section{Acknowledgment}
The BES collaboration thanks the staff of BEPC for their hard efforts
and the members of IHEP computing center for their helpful assistance,
and also T.P. Li for helpful discussion on 2D pdf function. 
This work is supported in part by the National Natural
Science Foundation of China under contracts
Nos. 19991480,10225524,10225525, the Chinese Academy of Sciences under
contract No. KJ 95T-03, the 100 Talents Program of CAS under Contract
Nos. U-11, U-24, U-25, and the Knowledge Innovation Project of CAS
under Contract Nos. U-602, U-34(IHEP); by the National Natural Science
Foundation of China under Contract No.10175060(USTC),and
No.10225522(Tsinghua University); and by the U.S. Department of Energy
under Contract No.DE-FG02-04ER41291 (U Hawaii).\par


\begin{thebibliography}{dd}

\bibitem{eichten} E.~Eichten {\em et al.}, Phys. Rev. D {\bf 21}, 203 (1980) and D {\bf 17},
3090 (1978); V.~A.~Novikov, {\em et al.}, Phys. Rep. {\bf 41}, 1 (1978); 
C.~Quigg and J.~L.~Rosner, Phys. Rep. {\bf 56}, 167 (1979);
G.~Bali, K.~Schilling and A.~Wachter, Phys. Rev. D {\bf 56}, 2566 (1997).

\bibitem{X3872} Belle, S.~K.~Choi {\em et al.}, Phys. Rev. Lett. {\bf 91}, 
262001 (2004). Babar, B. Aubert {\em et al.}, hep-ex/0406022.

\bibitem{etac and hc} Belle, S.~.K~Choi {\em et al.}, Phys. Rev. Lett. {\bf 89},
102001 (2002); CLEO, A.~Tomaradze, hep-ex/0410090; E835, C.~Patrignani,
hep-ex/0410085.

\bibitem{QCD review} N.~Brambilla {\em et al.}, Heavy Quarkonium Physics, hep-ph/0412158.

\bibitem{ebert}D.~Ebert, R.~N.~Faustov and V.~O.~Galkin, Phys. Rev. D67, 014027 (2003) 
and D {\bf 62}, 034014 (2000).

\bibitem{Radford} S.~F.~Radford and W~.W.~Repko, hep-ph/0409290, Sep 24, 2004.


\bibitem{artamonov} OLYA and MD-1, A.~S.~Artamonov {\em et al.}, 
Phys. Lett. B {\bf 474}, 427 (2000).

\bibitem{Aulchenko} KEDR, V.~M.~Aulchenko {\em et al.}, 
Phys. Lett. B {\bf573}, 63 (2003);
BES, J.Z. Bai et al., Phys. Lett. B {\bf 550}, 24 (2002);

\bibitem{pdg04} S. Eidelman {\em et al.} (Particle Data Group),
Phys. Lett. B {\bf 592}, 1 (2004).

\bibitem{bagnosco}
E835, S. Bagnosco {\em et al.}, Phys. Lett. B {\bf 533}, 237 (2002);
M.~Andreotti {\em et al.}, Phys. Rev. Lett. {\bf 91}, 091801 (2003).

\bibitem{CLEO chicJ} CLEO, S.B.~Athar {\em et al.},
Phys. Rev. D {\bf 60} 112002(2004). From their measured photon energies, 
one may estimate the $\chi_{cJ}$ masses using Eq.~(\ref{energyg}) in the text.

\bibitem{Godfrey and Rosner} S.~Godfrey and J.~L.~Rosner, 
Phys. Rev. D {\bf 66}, 014012(2002); S.~Godfrey, hep-ph/0501083.

\bibitem{bes2} BES, J.~Z.~Bai. {\em et al.}, Nucl. Instr. Meth.
{\bf A458}, 627 (2001); Nucl. Instr. Meth. {\bf A344}, 319 (1994).

\bibitem{liuhm} H.M.~Liu {\em et al.}, ``The BESII Detector Simulation'', 
to be submitted to NIM.

\bibitem{karl} G.~Karl, S.~Meshkov, and J.L.~Rosner, Phys. Rev. D {\bf 13}, 1203 (1976).

\bibitem{hongt} T.~Hong {\em et al.}, High Energy Phys. and Nucl. Phys.,
V {\bf 25}, 617 (2001).

\bibitem{mfisz} M.~Fisz, Probability and Mathematic Statistics,
Berlin, 1958.

\bibitem{mnfit} I.C.~Brock, A Fitting and Plotting Package 
Using MINUIT, version 4.07, Dec. 22th, 2000. 

\bibitem{Babar} For instance, see : Babar collab., Phys. Rev. D65, 091104(2002).

\bibitem{beam energy spread} The spread in the center-of mass energy at
$\psi(2S)$ energy is $1.3~\hbox{MeV}$. See also  BES Collab., 
Phys. Lett. B {\bf 550}, 24 (2002).

\bibitem{field correction} A correction factor $s = 0.9975\pm 0.0007$ 
for magnetic field at the BESII 
is determined with mass measurement of the $J/\psi$ reconstructed from 
$\psi(2S) \rightarrow \pi^+\pi^- J/\psi$ and $J/\psi \rightarrow \mu^+\mu^-$
decays.

\bibitem{chenjc} J.C.~Chen {\em et al.}, Phys. Rev. D {\bf 62}, 034003 (2000).

\end{thebibliography}
\end{document}